\documentclass[reprint,superscriptaddress,amsmath,amssymb,aps,prb,showpacs]{revtex4-1}

\usepackage[utf8]{inputenc}
\usepackage{graphicx}
\usepackage{amsmath}
\usepackage{amssymb}
\usepackage[exponent-product=\cdot,scientific-notation=false,separate-uncertainty=false,per-mode=symbol]{siunitx}
\usepackage[textsize=small]{todonotes}
\usepackage{booktabs}
\usepackage{color}

\begin{document}

\title{Reconstructing Detailed Line Profiles of Lamellar Gratings from GISAXS Patterns with a Maxwell Solver} 

% \cauthor[a]{Victor}{Soltwisch}{victor.soltwisch@ptb.de}{}
% \author[a]{Anal\'{i}a}{Fern\'{a}ndez Herrero}
% \author[a]{Mika}{Pflüger}
% \author[a]{Anton}{Haase}
% \author[b]{Jürgen}{Probst}
% \author[a]{Christian}{Laubis}
% \author[a]{Michael}{Krumrey}
% \author[a]{Frank}{Scholze}
% \aff[a]{Physikalisch-Technische Bundesanstalt (PTB), Abbestraße 2-12, 10587 Berlin, \country{Germany}}
% \aff[b]{Helmholtz-Zentrum Berlin (HZB), Albert-Einstein-Straße 15, 12489 Berlin, %\country{Germany}}
% \date{\today}

\author{V. Soltwisch}
\email{Victor.Soltwisch@ptb.de}
\affiliation{Physikalisch-Technische Bundesanstalt (PTB), 
Abbestr. 2-12, 10587 Berlin, Germany}

\author{A. Fern\'{a}ndez Herrero}
\affiliation{Physikalisch-Technische Bundesanstalt (PTB), 
Abbestr. 2-12, 10587 Berlin, Germany}

\author{M. Pflüger}
\affiliation{Physikalisch-Technische Bundesanstalt (PTB), 
Abbestr. 2-12, 10587 Berlin, Germany}

\author{A. Haase}
\affiliation{Physikalisch-Technische Bundesanstalt (PTB), 
Abbestr. 2-12, 10587 Berlin, Germany}

\author{J. Probst}
\affiliation{Helmholtz-Zentrum Berlin (HZB), Albert-Einstein-Str. 15,
12489 Berlin, Germany}

\author{C. Laubis}
\affiliation{Physikalisch-Technische Bundesanstalt (PTB), 
Abbestr. 2-12, 10587 Berlin, Germany}

\author{M. Krumrey}
\affiliation{Physikalisch-Technische Bundesanstalt (PTB), 
Abbestr. 2-12, 10587 Berlin, Germany}

\author{F. Scholze}
\affiliation{Physikalisch-Technische Bundesanstalt (PTB), 
Abbestr. 2-12, 10587 Berlin, Germany}

\date{\today}

\begin{abstract}
Laterally periodic nanostructures were investigated with grazing incidence small angle X-ray scattering (GISAXS) by using the diffraction patterns to reconstruct the surface shape. To model visible light scattering, rigorous calculations of the near and far field by numerically solving Maxwell's equations with a finite-element method are well established. The application of this technique to X-rays is still challenging, due to the discrepancy between incident wavelength and finite-element size. This drawback vanishes for GISAXS due to the small angles of incidence, the conical scattering geometry and the periodicity of the surface structures, which allows a rigorous computation of the diffraction efficiencies with sufficient numerical precision. 
To develop dimensional metrology tools based on GISAXS, lamellar gratings with line widths down to 55 nm were produced by state-of-the-art e-beam lithography and then etched into silicon. The high surface sensitivity of GISAXS in conjunction with a Maxwell solver allows a detailed reconstruction of the grating line shape also for thick, non-homogeneous substrates. 
The reconstructed geometrical line shape models are statistically validated by applying a Markov chain Monte Carlo (MCMC) sampling technique which reveals that GISAXS is able to reconstruct critical parameters like the widths of the lines with sub-nm uncertainty. 
 \end{abstract}

\maketitle
% \begin{synopsis}
% The shallow incidence angles in GISAXS allows to use a rigorous Maxwell solver in combination with the finite-element method for a reconstruction of nm sized periodic surface structures. 
% \end{synopsis}
%\fontsize{14}{12}\selectfont
\section{Introduction}
Measurements on the length scale of several nm are challenged by the atomic granularity of matter and by structures which cannot easily be described with simple models. The continuously shrinking patterns in the semiconductor industry are at the forefront of technological development regarding the requirements for size reproducibility and regularity.\\ 
Scanning probe microscopy techniques (e.g.~atomic force microscopy (AFM), scanning tunnelling microscopy (STM), scanning near-field optical microscopy (SNOM), scanning electron microscopy (SEM)) are powerful tools for the investigation of nanostructured surfaces and occupy the key positions for metrology tools in industry. However, in particular X-ray scattering is also an established technique in nanoscience. Grazing incidence small angle X-ray scattering (GISAXS)~\cite{guinier_small-angle_1955,levine_grazing-incidence_1989,renaud_probing_2009} also offers comparably fast measurements, in addition to being destruction-free.
With incidence angles close to the critical angle $\alpha_c$ of total external reflection, GISAXS is a technique with high surface sensitivity and is perfectly suited for in-situ applications. Due to the large photon beam footprint as compared to the nm pattern size, it directly yields statistical information on fluctuations such as structure roughness for a large structured area. 
However, the long beam footprint prevents probing specific small sample volumes in the illuminated area. A challenge for GISAXS is the characterization of structured surfaces with complex periodic and non-periodic sample layouts (e.g. photomasks). Changes in the design of the sample layout~\cite{Pfluger_2017a} allow to overcome this problem, but this results also in an extreme loss in scattering intensity which eliminates one of the major advantages of GISAXS for in-situ applications.
In contrast to GISAXS, transmission SAXS~\cite{wang_saxs_2007,Jones_2006,sunday_determining_2015} (CDSAXS) is able to probe much smaller sample volumes, only limited by the overall beam dimension. The drawback of the transmission geometry results from the fact that a significant portion of the incident beam may be absorbed in the substrate. As the incidence angle moves away from normal incidence, the beam path length increases and for typical silicon wafers X-ray photon energies above 13 keV are needed for sufficient transmission\cite{Sunday_2016spie}.
A combination of grazing incidence reflection and transmission measurements~\cite{Lu_2013} (GTSAXS) at the sample edges is often not practical because typical samples are not structured to the very edge. Therefore, GISAXS is ideally suited for the characterization of nano structured surfaces on non-homogeneous substrates.\\
%In contrast to SAXS in transmission geometry~\cite{wang_saxs_2007,Jones_2006,sunday_determining_2015}, GISAXS in reflection geometry also allows the study of structured surfaces of thick, non-homogeneous substrates.
Several groups have already performed GISAXS measurements on gratings for co-planar~\cite{tolan_x-ray_1995,metzger_nanometer_1997,jergel_structural_1999,mikulik_1999} and conical scattering geometry~\cite{mikulik_coplanar_2001,yan_intersection_2007}. The diffraction patterns are well understood. The intensity modulation of the scattered orders is related to the line shape (form factor) of the grating. To reconstruct the line shape parameters, including the line width, the sidewall angle or the line height, the so-called inverse problem has to be solved. The corresponding calculations for GISAXS were mostly done using the distorted wave Born approximation (DWBA) including an analytic expression of the form factor~\cite{Babonneau:hx5104,Lazzari:vi0158,renaud_probing_2009,rueda_grazing-incidence_2012,rauscher_small-angle_1995,jiang_waveguide-enhanced_2011-1,hofmann_grazing_2009,suh_characterization_2016,meier_situ_2012}.
Arbitrarily shaped structures must be discretized and require numerical solvers~\cite{Chourou:nb5076}. In the optical domain of the electromagnetic spectrum, modelling the light scattering by numerically solving the time-harmonic Maxwell's equations with a higher-order finite-element method~\cite{pomplun_adaptive_2007,Kato:12} is well established. If the periodic structures are invariant in one dimension (i.e.~along the grating lines), the computational domain can be reduced to two dimensions which  significantly decreases the computational effort and also allows the calculation of rather large domains as compared to the incident wavelength. This enables, together with the shallow incident angle in GISAXS, the application of this approach to short wavelengths as in X-ray scattering. Solving an inverse problem means minimizing the difference between the measured scatter intensities and the simulated intensities by adapting the geometrical parameters. A statistical validation of the optimized models is possible with the Markov chain Monte Carlo (MCMC) approach~\cite{emcee}, which gives the opportunity to obtain parameter sensitivities and delivers confidence intervals.

Besides the regular diffraction pattern from a line grating, the diffuse scattering background in the GISAXS pattern reveals additional information about the surface.
The appearance of long-range ordered superstructures in the scattering pattern yields information about the e-beam fabrication process.
Roughness and imperfections of the surface structures produce a complex diffuse scattering background, which can be described by kinematic and dynamic scattering effects. These effects are correlated with the grating line shape and also allow the determination of geometrical parameters~\cite{meier_situ_2012,Soltwisch2016}.  

\section{Experimental Details}

 \begin{figure} 
    \includegraphics[width=0.45\textwidth]{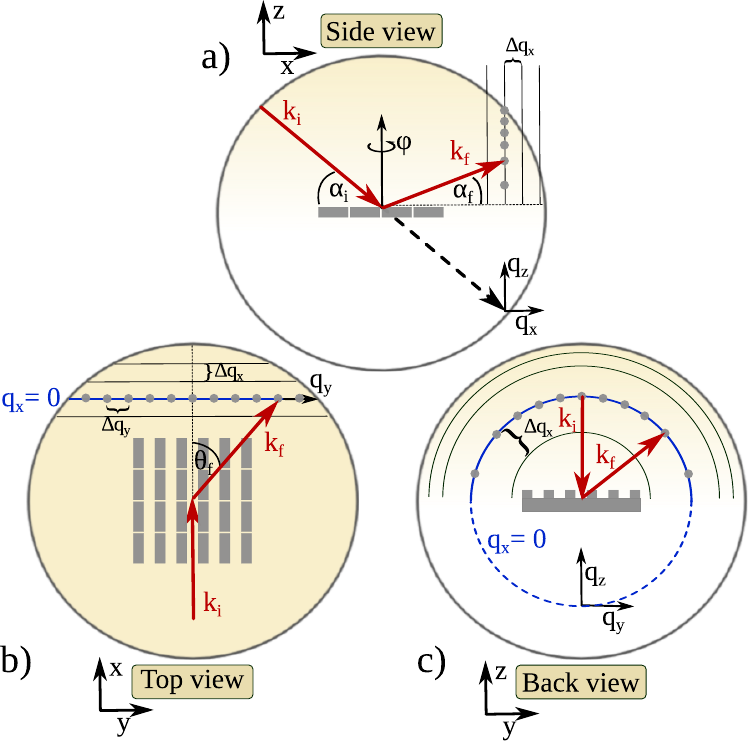}
\caption{Schematic views of the scattering geometry where the grating is orientated parallel to the incident beam. a) Side view where the scattered wave vector $k_f$ propagates along the exit angle $\alpha_f$.
b) Top view where the scattered wave vector $k_f$ includes the azimuthal exit angle $\theta_f$. The blue line visualizes the cut through the grating truncation rods from the grating structure at $q_x=0$ which leads to discrete diffraction orders (with $\Delta q_y$) on a semicircle on the detector. An additional periodicity (with $\Delta q_x$) in the lateral direction of the grating is visualized as black lines. This additional periodicity is also visible in the back view c). Note the non-periodic series of semicircles in the back view projection.}
\label{figure_sketch}
\end{figure}
The basic geometry of GISAXS is shown in Fig.~\ref{figure_sketch}. A monochromatic X-ray beam idealized as a plane wave with the wave vector $\vec{k}_i$ impinges on the sample surface at a grazing incidence angle $\alpha_i$. The elastically scattered wave with the wave vector $\vec{k}_f$ propagates along the exit angle $\alpha_f$ and the azimuthal angle $\theta_f$. Here we use the common notation for the scattering vector $\vec{q}=\vec{k}_f - \vec{k}_i$
\begin{equation}
	\left(\begin{array}{c} q_x \\ q_y \\ q_z \end{array}\right)=\left( \begin{array}{c} k_0 (\cos(\theta_f)\cos(\alpha_f)-\cos(\alpha_i)) \\
	 k_0 (\sin(\theta_f)\cos(\alpha_f)) \\
	  k_0 ( \sin(\alpha_f)+\sin(\alpha_i) )\end{array} \right)\\	
\label{eq:qvector}
\end{equation}
using angular coordinates (see Fig.~\ref{figure_sketch}) and $k_0 = 2\pi / \lambda$ with $\lambda$ as the wavelength.
Scattering from a periodically structured surface, e.g.~a grating, leads to a characteristic diffraction pattern~\cite{tolan_x-ray_1995,hofmann_grazing_2009,wernecke_direct_2012,rueda_grazing-incidence_2012,soccio_assessment_2014}.
The GISAXS pattern from a line grating observed with a 2D detector can be easily understood by a reciprocal space construction~\cite{mikulik_coplanar_2001,yan_intersection_2007}. The reciprocal space of a line grating, with pitch $p$, consists of truncation rods which are perpendicular to the surface and aligned with a spacing $\Delta q_y$ in the direction perpendicular to the scattering plane. The scattering pattern on the detector arises as a semicircle of discrete diffraction orders with $\Delta q_y = 2\pi/p$ which follows from the intersection of the Ewald sphere with a radius of $k_0$ and the reciprocal space representation of the lattice at $q_x=0$  (see Fig.~\ref{figure_sketch}).\\
\begin{figure}
\centering
\includegraphics[width=0.49\textwidth]{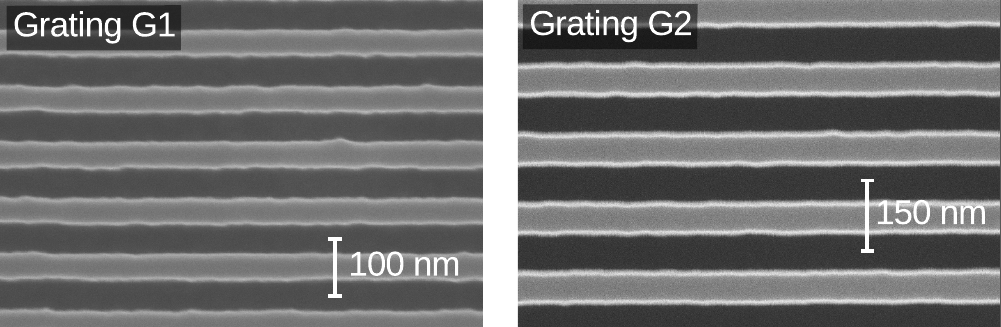}
\caption{Scanning electron microscopy (SEM) images of both lamellar gratings. Dark areas correspond to the grooves.}
\label{figure_sem}
\end{figure}
The GISAXS experiments were conducted at the four crystal monochromator beamline, operated by the Physikalisch-Technische Bundesanstalt (PTB), at the electron storage ring BESSY II in Berlin~\cite{krumrey_high-accuracy_2001}. The beamline offers a photon energy from $E=$ 1.75 keV to $E=$ 10 keV. By using a beam-defining 0.52 mm pinhole about 150 cm before the sample position and a scatter guard 1 mm pinhole about 10 cm before the sample, we reach a beam spot size of about 0.5 x 0.5 mm at the sample with minimal parasitic scattering. A 6-axis goniometer installed in a UHV chamber allows the alignment of the scattering angle $\alpha_i$ with an accuracy of $\pm 0.001^{\circ}$. The grating lines were aligned in parallel orientation ($\varphi =0^{\circ}$) with respect to the incident beam with an accuracy of $\pm 0.002^{\circ}$. An in-vacuum PILATUS 1M hybrid pixel detector is installed on a movable sledge which allows the sample-detector distance to be varied from 1.7 m up to 4.5 m \cite{wernecke_characterization_2014}. The detector features 20 bit counters for every pixel and consists of 10 separate modules. To extend the accessible photon energy range to the soft X-ray region (EUV), small angle X-ray scattering was also performed at the soft X-ray beamline~\cite{scholze_high-accuracy_2001}. An Andor CCD camera was mounted on a UHV chamber with a fixed sample detector distance of $0.75$ m, which allows the performance of extreme ultraviolet small angle scattering (EUV-SAS) at steeper incident angles $\alpha_i$ of $7^{\circ}$. \\

Two lamellar gratings were manufactured by electron beam lithography. Grating G1 with a nominal pitch of 100 nm and a line width of 55 nm and grating G2 with a nominal pitch of 150 nm and a line width of 65 nm (see Table \ref{table__gratings} and Fig.~\ref{figure_sem}). The grating areas measure 1 mm by 15 mm (grating G1) and 0.51 mm by 4 mm (grating G2) with the lines oriented parallel to the long edge. To manufacture the gratings, a silicon substrate was spin coated with the positive resist ZEP520A (organic polymer). Pattern generation was done using a Vistec EBPG5000+ e-beam writer, operated with an electron acceleration voltage of 100 kV~\cite{senn_fabrication_2011}. After resist development, the grating was etched into the silicon substrate via reactive ion etching, using the etching gases SF$_6$ and C$_4$F$_8$. Finally, the remaining resist was removed with oxygen plasma treatment.
Grating G1 was fabricated in 2013 and grating G2 in 2016 using a slightly different etching process. The resulting line shape is very sensitive to the conditions in the reactive ion etching chamber, which changed slightly due to other processes running in the same chamber between the manufacturing of the two samples, leading to the differences in geometry seen in Table \ref{table__gratings} and Fig.~\ref{figure_fem_sem}.

\section{Superstructures}

\begin{figure}
  \includegraphics[width=0.49\textwidth]{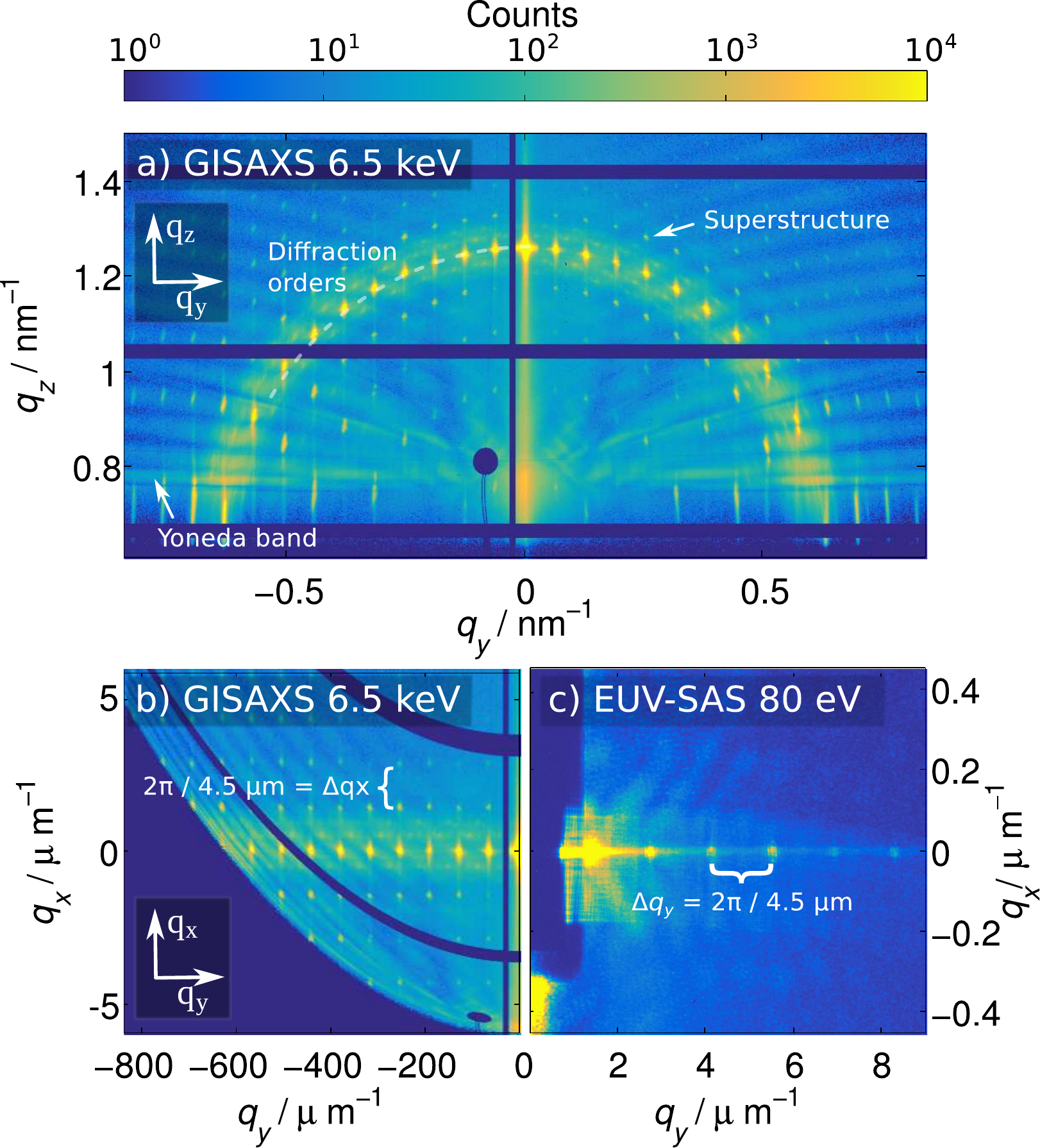}
\caption{a) GISAXS diffraction pattern of grating G1 obtained at $E= 6.5$ keV and $\alpha_i=1.1^{\circ}$  shown as a function of ($q_y,q_z$). b) The projection of the scatter intensity on the corresponding ($q_y,q_x$) map reveals that the superstructure is periodic in the lateral direction $x$. c) The identical periodic modulation is also visible in the $y$ direction by tuning the photon energy into the EUV range. The image was obtained at $E=80$ eV and $\alpha_i=7^{\circ}$ (EUV-SAS).}
\label{figure_qyqx}
\end{figure}
Besides the intense diffraction spots from the grating structure (see Fig. \ref{figure_qyqx}), the GISAXS pattern also shows a dominant superstructure in the ($q_y,q_z$) pattern as additional diffraction semicircles around the main grating diffraction semicircle at $q_x = 0$ (see Fig.~\ref{figure_qyqx} a)). The non-periodic shifting of the superstructures in the $q_z$ direction is clear evidence that the periodic modulation must be lateral along $q_x$. This behaviour is also illustrated in Fig.~\ref{figure_sketch}. A periodic modulation with the pitch $P$ of the grating in the lateral x direction leads to additional grating truncation planes with $\Delta q_x = 2\pi/P$. This periodicity appears in the detector plane as additional GISAXS diffraction semicircles with the same $\Delta q_y$ periodicity for the diffraction orders but including the $\Delta q_x$ spacing of the lateral modulation. For the evaluation of the lateral periodicity, the corresponding ($q_y,q_x$) projection must be calculated. With grazing incidence angles around 1$^{\circ}$ and a photon energy of 6.5 keV, the observable scattering vector component $q_x$ is in the order of $\mu \text{m}^{-1}$, corresponding to $\mu$m sized structures on the sample surface (cf. Fig. \ref{figure_qyqx} b)).
This allows the extraction of the lateral pitch size (along the grating lines) as  $\Delta x =$ 4.5 $\mu$m for the superstructure. This superstructure is directly related to the stitching field size of the e-beam writer (4.53$\times$ 4.53) $\mu \text{m}^2$. The lateral pitch size of the superstructure perpendicular to the scattering plane is not visible in the GISAXS scattering pattern, due to the different range for the $q_y$ scattering vector, but can be obtained with EUV-SAS at $\alpha_i$ = 7$^{\circ}$ and photon energies around $E = 80$ eV (see Fig.~ \ref{figure_qyqx} c)). 
Thus, the measurement demonstrates that small angle scattering techniques are also able to extract long-range ordering of nm structured surfaces due to the elongated beam footprint.
The finite-element approach in the X-ray spectral range, however, is limited to a two dimensional domain assuming a perfect infinite grating along the scattering direction. Such long-range perturbations can only be described in the model as a form of roughness.
\section{Computational diffraction intensities}
The reconstruction of the geometrical layout of nm sized structures by measuring the intensity distribution of the scattered photons means solving the inverse problem. In an iterative process, a theoretical model is used to compute the scattered intensity from a guess about the structure and compares the measured intensities with the simulation. 
In the last decade CDSAXS was developed as a powerful tool for the characterization of periodic nano structured surfaces. One major advantage of all transmission SAXS experiments, lies in the fact that a theoretical modelling with the first Born approximation (BA) is suitable for a surface shape reconstruction, as demonstrated by several groups\cite{Hu_2004,Wang_2009,sunday_determining_2015,Sunday_2016spie,Lu_2013}. The incoming wave scatters only once at the target before forming the scattered wave. With an analytic description of the line shape (form factor) this modelling approach is relatively fast.\\
With thick and non-homogeneous substrates the GISAXS measurement method is an interesting option for surface characterization. With decreasing incidence angle $\alpha_i$ the scattered amplitude increases while the transmission decreases. Incidence angles close to the external reflection allow to probe targets with a high surface sensitivity due to the comparably small penetration depths. This requires an extension of the BA, due to the fact that the BA neglects any multiple scattering effects.
For GISAXS the distorted wave Born approximation is well developed in several Software packages (IsGisaxs\cite{Lazzari:vi0158}, BornAgain\cite{durniak_gisas_2014}, HipGISAXS\cite{Chourou:nb5076}, FitGisaxs\cite{Babonneau:hx5104}) which are able to deal with second order multiple scattering effects. They extend the BA with additional scattering contributions from different multiple scattering events. The DWBA is able to explain a lot of GISAXS measurements, but higher order scattering effects (e.g. higher order Yoneda scattering\cite{Soltwisch2016}) are not covered in the typical implementation of the DWBA. The biggest advantages of both perturbation theories originate from the simplicity of the analytic form factor description and the resulting computational speed. This strong advantage reduces for arbitrary form factors, which must be discretized numerically\cite{Chourou:nb5076}. Beside those perturbation theories, solving the Maxwell's equation is also possible with a finite-element approach. Arbitrarily shaped objects can be implemented easily as vector functions and parametrized. In contrast to the approximation methods, a Maxwell solver allows the computation of real far field scattering intensities, including the solution of the near field in the computational domain. Higher order scattering effects can be easily studied with the computed electromagnetic field distributions inside the scatter objects. Additionally, the computation of local field distributions gives rise to the simulation of depth depended absorption measurements, e.g. grazing incidence X-ray fluorescence (GIXRF). Coupled with the strong increase in the available computational power in the last decade, solving the Maxwell's equations based on finite-elements is an interesting option in the X-ray spectral range. It is still limited to specific experimental settings (small angles, infinite gratings), i.e.~an expansion to CDSAXS is not applicable at the moment, but possible future applications are only limited by the computational power.\\ 
In the following we will give only a rather compact introduction on the topic of finite-element discretization. A detailed summary can be found elsewhere\cite{monk2003finite,pomplun_adaptive_2007}.\\  
\begin{figure}
\centering
\includegraphics[width=0.49\textwidth]{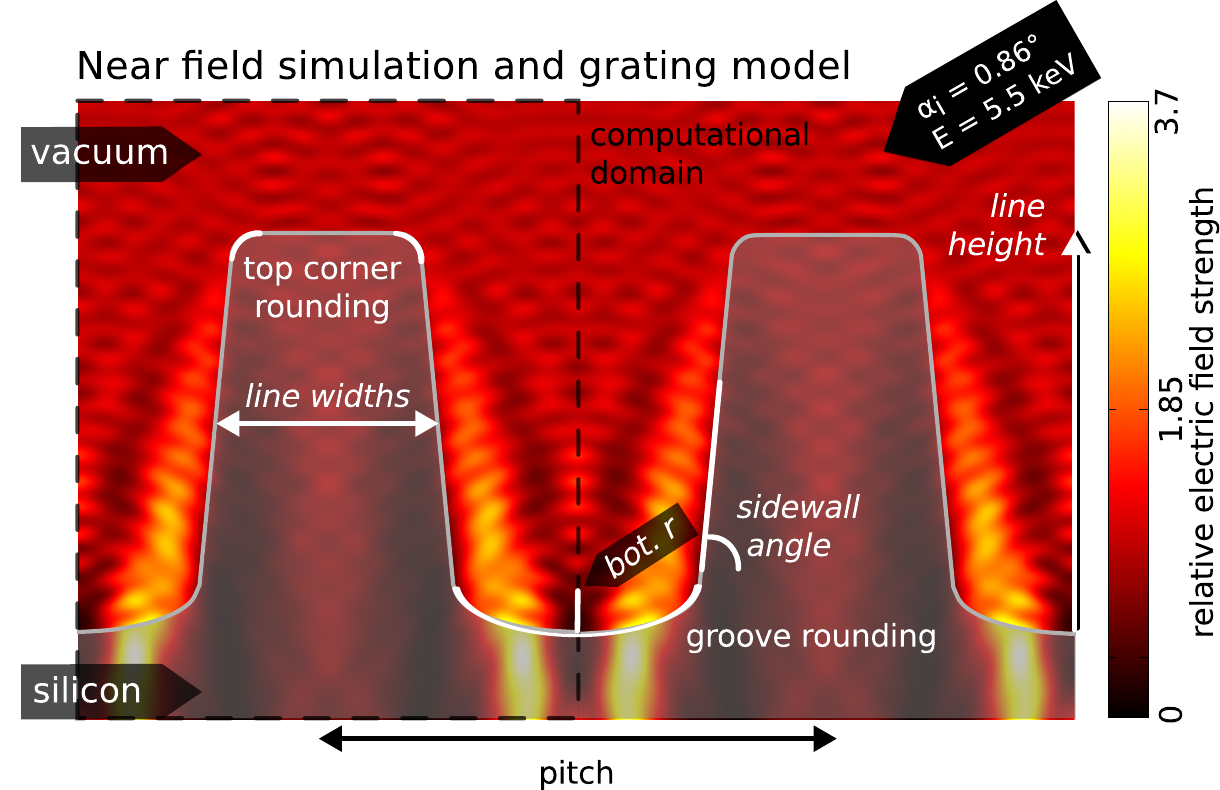}
\caption{Sketch of the computational domain and the grating model which was used for the construction of the finite-element mesh. An example of a near field simulation obtained with the Maxwell solver in a GISAXS geometry is shown in the background.}
\label{figure_sketch_fem}
\end{figure}

X-rays are treated as electromagnetic plane waves of 
wavelength $\lambda= h c_0 / E$ (with the Planck constant $h$, speed of light $c_0$ and photon energy $E$) which scatter on nano-structures. In this case, the set of Maxwell's equations can be rewritten to a single, second order curl-curl equation for the electric field\cite{pomplun_adaptive_2007}. 
%A variational formulation of the single equation is the basis for the finite-element discretization. It is obtained by multiplying with a vector valued test function and subsequent integration over $\mathbb{R}^3$. 
%The finite-element method thereby restricts the function space to a finite dimensional subspace. This means that the space, in which the solution is determined, will be approximated not the Maxwell equation itself. 
The general idea of the finite element discretization is that the computational domain is subdivided into small patches (e.g. triangles). On these patches a vectorial ansatz function is defined usually with polynomials with a fixed order $p$. The approximate electric field solution is the superposition of these local ansatz functions.
The numerical accuracy of the approximate electric field is a functional of the size of the finite-elements and the degree $p$ of the polynomials.

To compute diffraction efficiencies, we assume periodic structures 
which are invariant in one dimension (along the grating lines). 
Well-converged solutions are obtained with the package JCMsuite~\cite{pomplun_adaptive_2007} using a higher-order finite-element method.\\

We focus on a reconstruction of the form factor (the line shape) and not the structure factor (the grating pitch) which is well known from lithography~\cite{Buhr:2007}. Therefore, we fixed the pitch of the structures to the nominal values in the reconstruction process.
%That implies that the pitches of the structures were fixed to the nominal values in the reconstruction process.
In Fig.~\ref{figure_sketch_fem}, a computational domain and the corresponding near field calculation are shown for a typical GISAXS measurement geometry ($\alpha_i = 0.86^{\circ}$, $E = 5.5$ keV). The sketched line profile of the grating emphasizes the important structural parameters which were optimized in the reconstruction: the line width, the line height, the sidewall angle, the top corner rounding and the groove rounding. The line width of the structure is defined at the half-height. The parameter of the top corner rounding describes a circle with the respective radius. The bottom groove rounding is constructed with an elliptical shape. The major radius depends on the line width and the sidewall angle, the minor radius is parameterized for the reconstruction. Due to the reactive ion etching process, one could expect a strong groove rounding and only a minor rounding of the top corners because the top surface is still covered by the resist during etching (also visible in the SEM cross-section images, see Fig. \ref{figure_fem_sem}). This model allows an adequate description of the expected line shape, with a low number of parameters.

Roughness or imperfections of the lamellar gratings cannot be modelled directly with the finite-element approach due to the small discretization length required in the X-ray region. The computational domain size would go beyond any available computer memory. To account for the line edge or line width roughness damping effects on the diffraction intensities $I$ of higher diffraction orders, an analytic correction must be applied. The line edge and line width roughness (LER/LWR) of the gratings were taken into account, in the optimization process, with the well-known analytic approach \cite{mikulik_1999,kato_effect_2010} based on Debye-Waller damping 
\begin{equation}
I = I^\textrm{model} \cdot \exp\left(-{\sigma_r^2 q_{y}^2}\right)\text{.}
\label{eq:debyedamping}
\end{equation}
This allows the correction of the undisturbed computational diffraction intensities $I^\textrm{model}$ from the Maxwell solver in a post-process.
The damping factor $\sigma_r$ of the rms line roughness was also included in the optimization process.

\section{Uncertainty evaluation}

The grazing incidence conical diffraction and the invariance of the grating in the direction of the scattering plane result in a standing wave field with a much larger period than the wavelength (see Fig.~\ref{figure_sketch_fem}). This allows a significant increase in the size of the required discretization length $d$. It breaks the conventional rule of half the incident wavelength for the discretization to ensure numerical accuracy. This allows the use of a Maxwell solver based on the finite-element method to efficiently treat GISAXS applications.

\begin{figure}
  \includegraphics[width=0.46\textwidth]{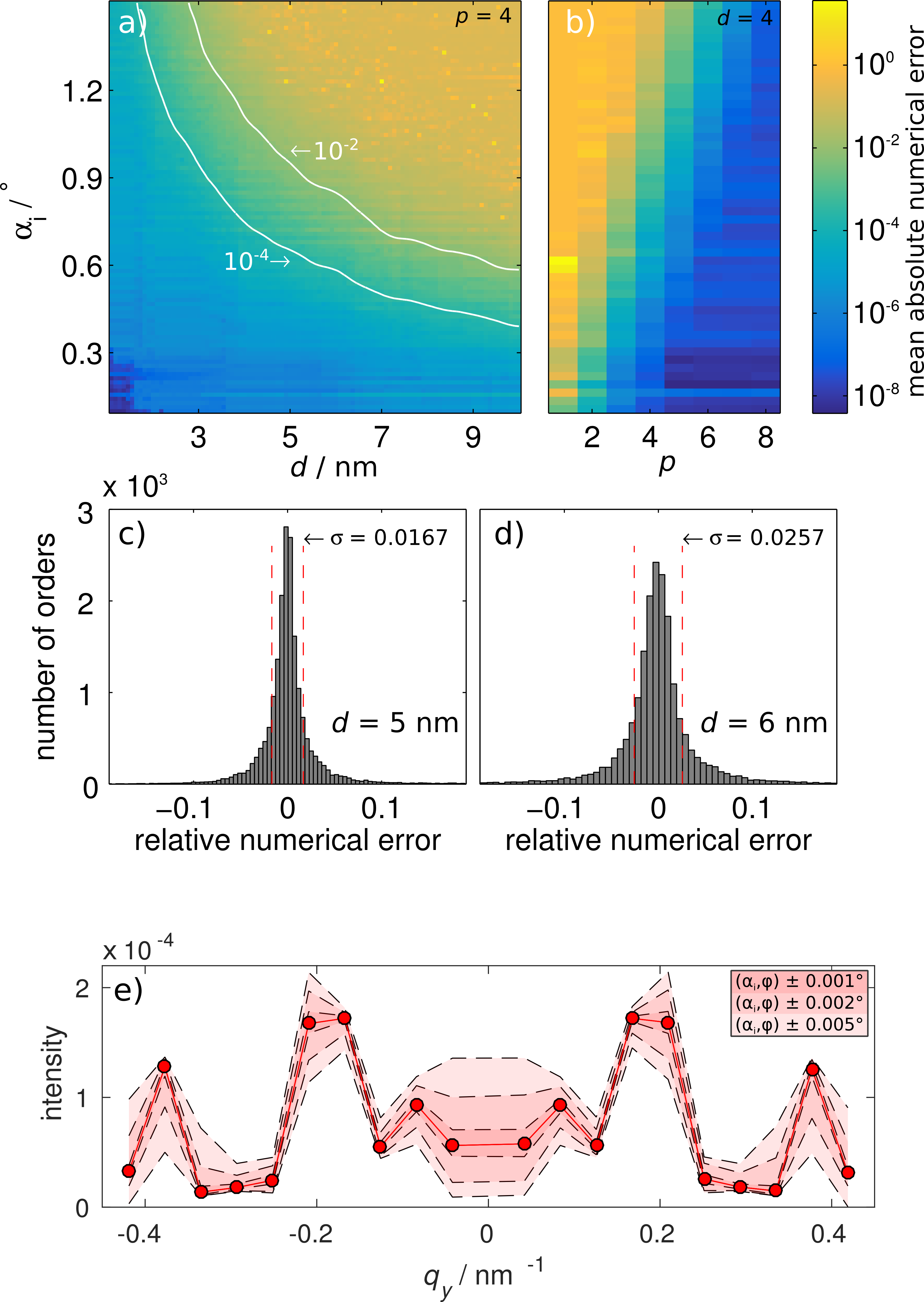}
\caption{The mean absolute numerical error (MANE) as a function of discretization length $d$  and polynomial degree $p$ of the ansatz function for different incidence angles $\alpha_i$. The polynomial degree $p$ was fixed to 4 in the simulation a) and the discretization length $d$ to 4 nm in the simulation b), calculated for a photon energy of 8 keV. Figures c) and d) compare the error distribution of all diffraction orders $N$ for a random distribution of different geometrical layouts for the reconstruction setup of grating G2.
e) Simulation of the sensitivities in the diffraction intensities for variations of the alignment angles ($\alpha_i,\varphi$) of grating G1.}
\label{figure_error_d_p}
\end{figure}

To ensure numerical accuracy with the finite angle of incidence, the relative numerical error of the simulated diffraction intensities (far field) was calculated for different numerical precision settings. The two main numerical degrees of freedom are the spatial discretization length $d$ and the polynomial degree $p$ which defines the ansatz function used to approximate the fields. The numerical errors are defined as the difference of the actual diffraction intensities $I^\textrm{model}$ to the quasi-exact results $I^\textrm{quasi}$. The quasi-exact calculation is defined as the converged computation with the highest achievable numerical precision settings, typically limited by the amount of available computational memory.
The numerical errors for typical GISAXS settings are shown in Figs.~\ref{figure_error_d_p} a) and b) for a photon energy of $E=8$ keV and varying grazing incidence angles $\alpha_i$. Similar results can be obtained for the azimuthal angle $\varphi$. This figure reveals the coupling of $d$ and $p$ to the incident angle $\alpha_i$ and allows the trade-off between the numerical precision and the computational effort to be estimated. For example, a decrease in the incident photon energy would shift the numerical precision in the figure towards larger incident angles. In this rough estimation, the size of the scatter objects represents a natural border of the discretization length $d$, but modern implementations of the finite-element method allow to make use of adaptive meshing algorithms\cite{Burger_2015} which are able to tune the local discretization length, e.g.~in critical regions like the vacuum-silicon interface. However, the obvious gap between the incident wavelength ($\lambda \ll 1$ nm) and the sufficient discretization length $d$ gives the opportunity for fast (below 1 s) and accurate simulations with a Maxwell solver based on the finite-element method. 
It should be noted that in Figs.~\ref{figure_error_d_p} a) and b), only the mean absolute numerical error (MANE) of all diffraction orders is visualized
\begin{equation}
\text{MANE} = \frac{1}{M}\bigg( \sum\limits_{N}\frac{|I^\textrm{quasi}_{N}-I^\textrm{model}_{N}|}{I^\textrm{quasi}_{N}}\bigg)\text{,}
\label{eq:relative_error}
\end{equation}
with $M$ being the number of diffraction orders and $N$ the specific diffraction order. This hides the fact that the numerical error for the different diffraction orders is coupled with the geometrical layout and the actual experimental settings. 
Diffraction orders with an exit angle $\alpha_f$, which is close to the critical angle of the substrate or the grating effective layer (Yoneda band)~\cite{Soltwisch2016}, are most affected. The number of usable diffraction orders is critical for the reconstruction process, as it directly limits the possible complexity of the reconstructed model. To achieve an upper estimation of the numerical errors, we simulated 1000 randomly distributed geometrical grating layouts within the parameter boundaries for the extraction of the quasi-exact diffraction intensities. The identical grating layouts were simulated with the numerical discretization which will later be used in the specific grating reconstruction process (see Table \ref{table__gratings}). We then compared every diffraction intensity with the corresponding quasi-exact computation. 
The results of this numerical error estimation for the reconstruction of grating G2 are exemplified visually in 
Figs.~\ref{figure_error_d_p} c) and d) for discretization lengths of $5$ nm and $6$ nm.
The dashed red lines mark the $\sigma$ interval. These $\sigma$ values in the range of several percent reveal that numerical precision has a significant impact in the uncertainty evaluation.
To achieve a better reconstruction result, the highest diffraction orders  should be neglected in the fitting process due to the numerical instabilities at the critical angle (c.f.~Fig.~\ref{figure_qyqx} a) Yoneda band). We also excluded the diffraction orders which are not used in the subsequent reconstruction process (e.g.~zero order) of grating G1 and G2.\\

Besides these numerical errors, small angle X-ray scattering is challenging for every experimental setup, especially if the accuracy of the alignment angles $\alpha_i$ and $\varphi$ is important. Under grazing incidence, small variations in the angles of incidence could change the measured diffraction efficiencies rapidly. This angular sensitivity is demonstrated in Fig. \ref{figure_error_d_p} e). The red dots are the simulated diffraction intensities (quasi-exact) of a typical lamellar silicon grating (grating G1) at $\alpha_i=0.7^{\circ}$ and $E=8$ keV. The changes in the diffraction intensities for angular variations in the order of $10^{-3}$ degrees are visualized with dashed lines. These angular uncertainties are close to the achievable motor step resolution in our experimental chamber. This is in principle not a big issue for the reconstruction process, because a subsequent calibration of the angles is still possible. But this calibration is also limited by the uncertainty from the estimation of the pixel size for the PILATUS detector~\cite{wernecke_characterization_2014} and leads to a very similar uncertainty of $0.002^{\circ}$.
To ensure an accurate reconstruction of the line shapes, both incidence angles must be included in the optimization process as independent parameters. This angular sensitivity in grazing incidence also influences the computational effort which is needed for an accurate simulation of measured diffraction intensities due to the divergence of the incident beam. 
The horizontal divergence of $0.01^{\circ}$ leads to an elongation of the diffraction peaks along the $q_z$ axis. The impact is directly visible in Fig.~\ref{figure_qyqx} a) for diffraction orders close to the horizon. To account for this angular distribution in the theoretical evaluation, we calculated 5 azimuthal incidence angles $\varphi$ for every simulation. The extracted diffraction intensities were weighted with a Gaussian distribution and the standard deviation of the incident beam divergence. Convergence studies with increasing numbers of incidence angles $\varphi$ (up to 50) reveal that the relative uncertainty $\sigma_{\text{div}}$ of the sampling with only 5 angles is in the region of 8\%.  A further reduction of this uncertainty is possible but would result in a significantly increased computational effort.\\

By using a photon counting detector, which follows Poisson statistics, for the measurements of diffraction intensities, the statistical measurement uncertainty $\sigma_{\text{N}}$ is typically an order of magnitude below the numerical uncertainty of the Maxwell solver. However, the homogeneity of the PILATUS detector at incident photon energies around 5 keV~\cite{wernecke_characterization_2014} is not negligible, which leads to an additional uncertainty $\sigma_{\text{hom}}$ of 2.5\% in the measured diffraction intensities.   
The distribution of the incident photon energies were taken into account, 
with the implementation of a Gaussian prior, with the bandwidth of $10^{-4}$ for each photon energy. For hard X-rays, the influence of uncertainties from the optical constants has almost vanished, especially for silicon far away (several keV) from any absorption edge. In the soft X-ray or EUV region, the situation could change dramatically and must be evaluated separately.

\section{Reconstruction of the line shape}

The general problem in a reconstruction process is the question of how much information is necessary to obtain a univocal solution. The constraints for the numbers of diffraction orders for a specific incident angle and photon energy are dictated by the grating pitch. This strictly limits the accessible information and can only be compensated for by mapping different regions of the reciprocal space (e.g.~tuning the incidence angles or photon energy). To avoid the experimental uncertainties during angular scans ($\alpha_i,\varphi$), we utilize the high stability of the four crystal monochromator and tune only the incident photon energy.
The intensity of the specular reflection is often disturbed by the mismatch between the beam spot size and the grating target size. Parts of the beam are reflected from the surrounding substrate and not from the structured surface. This distorts the zero order diffraction intensity and prevents the extraction of absolute intensities. Fitting the relative diffraction intensities of the non-zero orders is thus more accurate for sample sizes below the size of the elongated beam footprint. 

To compare the measured diffraction intensities with theoretical values, the Maxwell solver computes the near field solution for a specific parameter set depending on the model and extracts the theoretical photon flux for every diffraction order by post-processing using a Fourier transformation. The minimization functional of the optimization problem $\chi^2$ is defined as the total sum of the least-squares functionals for every photon energy $E$,
\begin{equation}
\chi^2 = \sum\limits_{E} \tilde{\chi}^2(\text{E})\text{,} 
\label{eqn:total_chi_2}
\end{equation}
where each of the functionals is defined as
\begin{equation}
\tilde{\chi}^2(E) = \sum\limits_{m} \frac{\big( I_m^\text{model}(E) 
- I_m^\text{meas}(E)\big)^2}{{\sigma}^2(E)}  \text{.} 
\label{eqn:chi_2}
\end{equation}
For the standard deviation $\sigma (E)$, the individual experimental and numerical uncertainty is propagated according to the Gaussian error propagation law,
\begin{equation}
{\sigma}^2(E) = \sigma_{\text{num}}^2(E) + \sigma_{\text{N}}^2 + \sigma_{\text{hom}}^2 + \sigma_{\text{div}}^2 \text{.} 
\label{eqn:sigma}
\end{equation}
The numerical uncertainties $\sigma_{\text{num}}(E)$ were precalculated for the different photon energies which were used in the measurements and for both gratings, as described in the previous section. The discretization length $d$ and the polynomial degree $p$ were chosen such that the numerical uncertainties were $\sigma_{\text{num}}(E) \approx 2.6\%$ (grating G2) and $ \sigma_{\text{num}}(E) \approx  3.6\%$ (grating G1). \\

A heuristic optimization method is ideally suited to explore a large parameter space of the inverse problem, but requires massive parallelization to reduce the computational effort to a reasonable time. With a well-optimized Maxwell solver and a commercially available workstation, we were able to solve 10$^6$ structures in less than one day. We used the particle swarm optimization~\cite{kennedy_particle_1995} method, which ideally delivers the global minimum of the total $\chi^2$ functional with an acceptable computational effort. However, no information about parameter sensitivity or the quantification of confidence intervals is delivered with a particle swarm optimization. To solve this issue, we applied an affine invariant Markov chain Monte Carlo sampling technique~\cite{emcee}. The likelihood is given by
\begin{equation}
P(\vec{x}) \propto \exp \big(-{\chi}^2 / 2 \big) \text{,} 
\label{eqn:likelihood}
\end{equation}
where $\vec{x}$ is the set of parameters of the model. As a starting point, a random set of parameters was generated with respect to predefined boundaries around the global minimum of the particle swarm optimization.
The confidence intervals and mean values were calculated by evaluating the 
probability distribution for each parameter, after the MCMC procedure converged. The confidence intervals ($\pm 1 \sigma$) given here represent 
percentiles of the number of samples found in the interval defined by the upper 
and lower bounds.\\

\begin{figure*}
\includegraphics[width=1\textwidth]{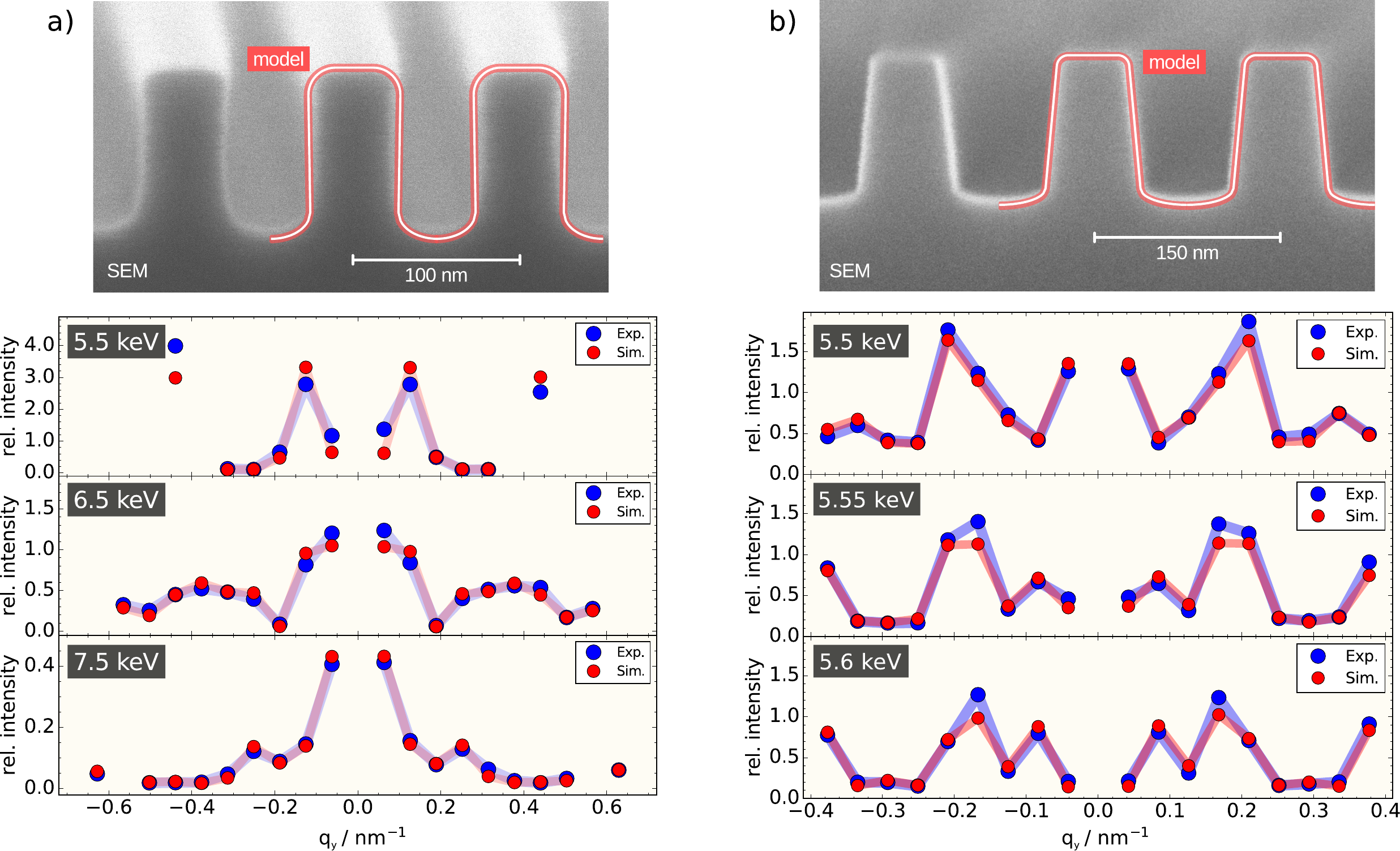}
	\caption{a) Cross-section SEM image and extracted diffraction efficiencies obtained at 5.5 keV up to 7.5 keV and $\alpha_i = 1.09^\circ$ from grating G1.
b) Cross-section SEM image and extracted diffraction efficiencies obtained at 5.5 keV up to 5.6 keV and $\alpha_i = 0.86^\circ$ from grating G2.    
The extracted diffraction efficiencies (blue dots) in both figures are shown as a function of the scattering vector $q_y$ (with $q_x$ = 0). The red dots represent the simulated diffraction efficiencies optimized with the MCMC sampling technique. The 0$^{th}$ order was overexposed and removed from the optimization process. The red lines in the cross-section SEM images obtained from witness samples illustrate the reconstructed line profile.}
\label{figure_fem_sem}
\end{figure*}
The results of the MCMC optimization are summarized in Fig.~\ref{figure_fem_sem} and Table \ref{table__gratings}. In Fig.~\ref{figure_fem_sem}, the extracted diffraction intensities for both gratings are shown as a function of the scattering vector component $q_y$ and photon energy, and compared to the simulated diffraction intensities of the MCMC optimization. Diffraction orders which were influenced by the detector gaps (dark regions in the GISAXS images, see Fig.~\ref{figure_qyqx}) were manually taken out of the reconstruction (missing points in Fig.~\ref{figure_fem_sem}). The good agreement between the measurement and simulation is evidence that the chosen model is able to describe the diffraction patterns from the grating structures. Also, the reconstructed geometry fits well to the expected nominal values. This is illustrated in Fig.~\ref{figure_fem_sem} by a comparison of the reconstructed line shape (c.f.~red line) with cross-section SEM images which we obtained from witness samples. 
A direct comparison depends on the homogeneity of the grating sample, because the SEM images show only a rather small part of the grating in contrast to the GISAXS measurements, which capture the whole structured surface. However, the reconstructed grooves show a significant rounding which is expected from the masked ion beam etching process. This leads to well-defined corners on the top and more rounded structures inside the grooves. This behaviour corresponds well with the reconstructed line shapes of both gratings.\\

A comparison of the reconstructed parameter values of both gratings and the estimated confidence intervals can be found in Table \ref{table__gratings}.
%{\renewcommand{\arraystretch}{1.2}
\begin{table}
\centering
\caption{The optimization results of the geometrical model parameters obtained from gratings G1 and G2 with the MCMC sampling, with corresponding confidence intervals ($\pm 1 \sigma$).}
\label{table__gratings}
\begin{tabular}{@{}lrcrc@{}}

%\toprule
\hline
\textrm{Parameter}& \textrm{Grating G1}& \textrm{Limits}& \textrm{Grating G2}& \textrm{Limits}\\
%\midrule
\hline
\textrm{pitch / nm} & 100 & \textrm{fixed} & 150 & \textrm{fixed}\\
\textrm{line height / nm} & 102.71 $\pm 0.12$ & [95 105] & 119.50 $\pm 0.11$ &[115 125]\\  
\textrm{line width / nm} & 54.04 $\pm 0.56$ & [50 60] & 67.30 $\pm 0.31$& [60 70]\\ 
\textrm{top r / nm} &  16.27 $\pm 0.39$ & [0 20] & 9.16 $\pm 0.39$ & [0 20]\\
\textrm{bot. r / nm} &  15.79 $\pm 0.24$ & [0 30] & 13.02 $\pm 0.58$ & [0 30]\\
\textrm{sidewall angle / }$^{\circ}$ & 90.91 $\pm 0.38$ & [80 95] &  84.73 $\pm 0.33$ & [80 90]\\
$\sigma_r$ \textrm{(rms) / nm} &  1.35 $\pm 0.07$ & [0 5] & 1.87 $\pm 0.14$ & [0 5]\\
%\bottomrule
\hline
\end{tabular}
\end{table}

The confidence intervals of the height are rather small with $\pm 120$ pm as compared to expected process variations of the etching. One could argue that the estimated confidence intervals correspond to the well-defined mean value of the grating line height over the large illuminated beam footprint area. The measured diffraction intensities can be understood as superposition of the diffraction patterns from several different gratings with slightly different parameters due to stochastic roughness or other systematic deviations in the etching process. This effect was accounted for by the inclusion of the Debye-Waller damping in our model. It should, however, be mentioned that this Debye-Waller damping approach may not hold for any kind of process variation as it was derived for line edge roughness in a binary grating model. In particular, the line height roughness does not disturb the two-dimensional periodicity of the surface, as line edge roughness does, and might have a different impact. The estimated confidence intervals are therefore only valid within the framework of the model presented here. For a complete uncertainty evaluation in GISAXS, the impact of these modelling assumptions must be further investigated.

\section{Summary}
We investigated lamellar gratings etched in silicon with pitch sizes of 100 nm and 150 nm using GISAXS in conical orientation.
The very distinct diffraction patterns from the rather similar grating line shapes highlight the high sensitivity of GISAXS to structure details. A dominant superstructure in the diffraction patterns is correlated to artefacts of the e-beam writing process and may in future be exploited to further tune the lithography process.
The conical diffraction geometry combined with grazing incidence angles allows the use of the finite-element approach with a Maxwell solver in the X-ray spectral range.
The Maxwell solver in conjunction with the finite-element method is a versatile tool to solve the inverse problem for highly periodic structures because it allows arbitrary geometric models of the line shape to be parameterized. A detailed reconstruction of the line shape is possible due to a significant reduction of the computational effort by adapting the discretization length to the grazing angle of incidence. It reveals geometric details, such as strong rounding inside the grooves, which are often not accessible with other direct non-destructive measurement methods like AFM. A first parameter sensitivity evaluation based on the Markov chain Monte Carlo method is presented. 
\section{Acknowledgment}
We thank the European Metrology Research Programme (EMRP) for financial support within the Joint Research Project IND17 "Scatterometry". The EMRP 
is jointly funded by the EMRP participating countries within EURAMET and the European Union.

%\referencelist[reference]
\bibliographystyle{apsrev4-1}
\bibliography{reference}
\end{document}